\documentstyle[psfig]{mn}

\newif\ifAMStwofonts
\ifoldfss
  \ifCUPmtlplainloaded \else
    \NewTextAlphabet{textbfit} {cmbxti10} {}
    \NewTextAlphabet{textbfss} {cmssbx10} {}
    \NewMathAlphabet{mathbfit} {cmbxti10} {} % for math mode
    \NewMathAlphabet{mathbfss} {cmssbx10} {} %  "   "    "
  \fi
  \ifAMStwofonts
    \ifCUPmtlplainloaded \else
      \NewSymbolFont{upmath} {eurm10}
      \NewSymbolFont{AMSa} {msam10}
      \NewMathSymbol{\upi}     {0}{upmath}{19}
      \NewMathSymbol{\umu}     {0}{upmath}{16}
      \NewMathSymbol{\upartial}{0}{upmath}{40}
      \NewMathSymbol{\leqslant}{3}{AMSa}{36}
      \NewMathSymbol{\geqslant}{3}{AMSa}{3E}

       \let\le=\leqslant
       
    \fi
  \fi
\fi % End of OFSS

\ifnfssone
  \newmathalphabet{\mathit}
  \addtoversion{normal}{\mathit}{cmr}{m}{it}
  \addtoversion{bold}{\mathit}{cmr}{bx}{it}
  \newmathalphabet{\mathbfit} % math mode version of \textbfit{..}
  \addtoversion{normal}{\mathbfit}{cmr}{bx}{it}
  \addtoversion{bold}{\mathbfit}{cmr}{bx}{it}
  \newmathalphabet{\mathbfss} % math mode version of \textbfss{..}
  \addtoversion{normal}{\mathbfss}{cmss}{bx}{n}
  \addtoversion{bold}{\mathbfss}{cmss}{bx}{n}
  \ifAMStwofonts
    \ifCUPmtlplainloaded \else
      \UseAMStwoboldmath
      \makeatletter
      \new@mathgroup\upmath@group
      \define@mathgroup\mv@normal\upmath@group{eur}{m}{n}
      \define@mathgroup\mv@bold\upmath@group{eur}{b}{n}
      \edef\UPM{\hexnumber\upmath@group}
      \new@mathgroup\amsa@group
      \define@mathgroup\mv@normal\amsa@group{msa}{m}{n}
      \define@mathgroup\mv@bold\amsa@group{msa}{m}{n}
      \edef\AMSa{\hexnumber\amsa@group}
      \makeatother
      \mathchardef\upi="0\UPM19
      \mathchardef\umu="0\UPM16
      \mathchardef\upartial="0\UPM40
      \mathchardef\leqslant="3\AMSa36
      \mathchardef\geqslant="3\AMSa3E

       \let\le=\leqslant

    \fi
  \fi
\fi % End of NFSS release 1

\ifnfsstwo
  \DeclareMathAlphabet{\mathbfit}{OT1}{cmr}{bx}{it}
  \SetMathAlphabet\mathbfit{bold}{OT1}{cmr}{bx}{it}
  \DeclareMathAlphabet{\mathbfss}{OT1}{cmss}{bx}{n}
  \SetMathAlphabet\mathbfss{bold}{OT1}{cmss}{bx}{n}
  \ifAMStwofonts
    \ifCUPmtlplainloaded \else
      \DeclareSymbolFont{UPM}{U}{eur}{m}{n}
      \SetSymbolFont{UPM}{bold}{U}{eur}{b}{n}
      \DeclareSymbolFont{AMSa}{U}{msa}{m}{n}
      \DeclareMathSymbol{\upi}{0}{UPM}{"19}
      \DeclareMathSymbol{\umu}{0}{UPM}{"16}
      \DeclareMathSymbol{\upartial}{0}{UPM}{"40}
      \DeclareMathSymbol{\leqslant}{3}{AMSa}{"36}
      \DeclareMathSymbol{\geqslant}{3}{AMSa}{"3E}

       \let\le=\leqslant

    \fi
  \fi
\fi % End of NFSS release 2

\ifCUPmtlplainloaded \else
  \ifAMStwofonts \else % If no AMS fonts
    \def\upi{\pi}
    \def\umu{\mu}
    \def\upartial{\partial}
  \fi
\fi

\title[UCDs in clusters of galaxies]{Origin of 
lower velocity dispersions  of ultra-compact 
dwarf galaxy populations in clusters of galaxies}
\author[K. Bekki]
       {K. Bekki \\
        School of Physics, University of New South Wales, 
Sydney 2052, NSW, Australia}
\date{Accepted 
      Received
      in original form 2001}

\pagerange{\pageref{firstpage}--\pageref{lastpage}}
\pubyear{1994}

\begin{document}

\maketitle

\label{firstpage}

\begin{abstract}

Recent observations have revealed that velocity dispersions
of ``ultra-compact dwarf'' (UCD) galaxies are significantly smaller
than those of other  galaxy populations 
in the Fornax and the Virgo clusters of galaxies. 
In order to understand the origin of the observed lower velocity
dispersions of UCDs, we numerically investigate 
line-of-sight velocity dispersion (${\sigma}_{\rm los}$)
of galaxy populations with variously different orbits 
in clusters of galaxies with the total masses of $M_{\rm cl}$.
We particularly investigate radial velocity dispersion
profiles (${\sigma}_{\rm los}(R)$) and  velocity dispersions
within the central 200 kpc of a cluster model  (${\sigma}_{m}$) for galaxies
with different pericenter distances ($r_{\rm p}$)
and orbital eccentricities ($e$)  
in the model  with $M_{\rm cl} = 7.0 \times 10^{13} {\rm M}_{\odot}$ 
reasonable for the Fornax cluster.
We find that ${\sigma}_{\rm los}(R)$
and ${\sigma}_{m}$ of galaxies with smaller $r_{\rm p}$
are steeper and smaller, respectively,
for a given initial $e$ distribution of galaxies.
For example, we find that 
${\sigma}_{m}$  is $\sim 260$ km s$^{-1}$ for
galaxies with $r_{\rm p} <50$ kpc and $\sim 336$ km s$^{-1}$  
for all galaxies in the model with the mean $e$ of 0.6.
These results imply that the observed lower
velocity dispersion of UCD population
is consistent with the UCDs having
significantly smaller $r_{\rm p}$ than other galaxy populations
in the Fornax.
We discuss these results in the context of the ``galaxy threshing''
scenario in which UCDs originate from nuclei of nucleated dwarf
galaxies.
We suggest that the observed differences in kinematical properties
between UCDs and other dwarf galaxy populations 
in clusters of galaxies can be understood
in terms of the differences in orbital properties 
between UCDs and the dwarf populations.

\end{abstract}


\section{Introduction}

Since very compact and  luminous stellar systems 
were  discovered in the central region
of the Fornax cluster of galaxies
(e.g, Hilker et al. 1999;  Drinkwater et al. 2000a, b),
physical properties of these systems -- now referred to as ``ultra-compact
dwarf'' (UCD) galaxies -- have been extensively investigated
by  observational studies (e.g., Phillipps et al. 2001;
Mieske et al. 2002, 2004, 2006;
Drinkwater et al. 2003; 
Ha\c segan et al. 2005;
Karick et al. 2006; Firth et al. 2006).
These observations have reported very 
unique properties of UCDs,  such as very compact sizes
($<100$ pc),  possibly higher mass-to-light-ratio ($M/L$) 
indicative of the presence of dark matter,
and scaling relations of dynamical 
properties (e.g., internal velocity dispersions) different from those of GCs
(e.g., Drinkwater et al. 2003; Ha\c segan et al. 2005).
Physical properties  of UCDs in different clusters and groups
of galaxies are now being investigated  (e.g., Jones et al. 2006;
Kilborn et al. 2005).

In spite of these observational progresses, it remains
unclear how UCDs formed and evolved  in the central regions of clusters
of galaxies (e.g.,  Bekki et al. 2001, 2003a;
Fellhauer \&  Kroupa 2002).
Bekki et al. (2001, 2003a) proposed  the ``galaxy threshing'' scenario
in which UCDs originate from  nuclei of nucleated dwarf galaxies
that had been destroyed by strong tidal fields of clusters of galaxies.
Fellhauer \&  Kroupa  (2002) proposed that numerous smaller young
star clusters, such as those observed in actively star-forming
Antennae galaxy, can finally evolve into a single UCD after
merging of them.
Although the observed structural properties of UCDs 
(e.g., De Propris et al. 2005),  scaling relations of dynamical
properties of UCDs (e.g., Evstigneeva et al. 2007),
and dynamical masses including possible dark matter halos
(Hilker et al. 2007)
have given some constrains on the above theoretical models,
it has not yet been determined which model can be regarded
as more reasonable.

Recent spectroscopic observations have reported that
the line-of-sight velocity dispersion (${\sigma}_{\rm los}$)
of UCDs (${\sigma}_{\rm los} = 246 \pm 25$ km s$^{-1}$)
is significantly smaller than those of other dwarf galaxy
populations ($396 \pm 41$  km s$^{-1}$)
in the Fornax cluster of galaxies (Gregg et al. 2007).
They also reported  that  ${\sigma}_{\rm los}$ of UCDs
is even smaller than that of GCs 
(${\sigma}_{\rm los} = 334 \pm 11$ km s$^{-1}$)
around NGC 1399 derived by Dirsch et al. (2004).
Mieske et al. (2004) have also 
reported that the bright compact objects in the Fornax cluster
shows a lower velocity dispersion than the dwarf galaxy population. 
The lower velocity dispersion has been also observed in
the six UCDs of the Virgo cluster of galaxies (Jones et al. 2006).
The origin of the observed lower of ${\sigma}_{\rm los}$ 
of UCDs in these two clusters
is one of important unresolved problems related to the origin of 
UCDs.

The purpose of this paper is thus to try to explain the observed
lower velocity dispersions  of UCDs based on somewhat  
idealized orbital evolution models of galaxies in clusters of galaxies.
The galaxy threshing scenario suggested that nucleated dwarf
galaxies with small pericenter
distances ($r_{\rm p}$) 
can become UCDs: galaxy populations with only a limited range of
orbital properties can be progenitors of UCDs (Bekki et al. 2001; 2003a).
We therefore investigate how radial profiles of  ${\sigma}_{\rm los}$
(${\sigma}_{\rm los}(R)$) and velocity dispersions within 
the central 200 kpc (${\sigma}_{\rm m}$) for galaxies in a cluster
depend on orbital properties of the galaxies (e.g., more circular or
eccentric orbits) and thereby  find galaxy populations that
show a lower velocity dispersion in the cluster.

The plan of the paper is as follows: in the next section,
we describe the orbital evolution models of galaxies in
a cluster of galaxies.
In \S 3, we present the results  
on   ${\sigma}_{\rm m}$  and ${\sigma}_{\rm los}(R)$ 
for galaxies with different orbital properties.
\S 4, we discuss the origin of flattened spatial distribution
of UCDs in the Fornax cluster of galaxies based on our previous
numerical simulations.
We summarise our  conclusions in \S 5.

\section{The model}

\subsection{The threshing scenario for UCD formation}

Although there could be a number of explanations for
the origin of the observed lower velocity dispersions
of UCDs in clusters of galaxies,
we here focus exclusively on the threshing scenario 
and thereby discuss the observation.
The predictions of the threshing scenario are summarized as follows:
(i) dE,Ns with smaller pericenter
distances ($r_{\rm p}$) and higher orbital eccentricities ($e$)
are more likely to be transformed into UCDs,
(ii) dE,Ns in the Fornax cluster 
can be transformed into UCDs, if the pericenter 
distances ($r_{\rm p}$)
of their orbits (with respect to the center of the cluster)
are smaller than the ``threshing radius'' ($R_{\rm th}$),
(iii) $R_{\rm th}$ is typically $\sim$ 50 kpc for dE,Ns
that have B-band magnitudes ($M_{\rm B}$) of $\approx -16$ mag
and are embedded in dark matter halos with large cores
(see Figure7 in Bekki et al. 2003a),
and (iv)  $R_{\rm th}$ depends on $M_{\rm B}$
for a given cluster mass
in such a way that $R_{\rm th}$ is smaller  for
larger (i.e., fainter)  $M_{\rm B}$.

The proposed threshing radius has been discussed
in the context of the observed compact spatial distribution
of UCDs in the Fornax cluster (Bekki et al. 2003a).
The orbital properties of the UCDs transformed from
dE,Ns with $r_{\rm p} < R_{\rm th}$ however have not been
discussed at all even in a qualitative way.
The threshing scenario implies  that orbital properties
might well be  different between UCDs (``destroyed populations'')
and the  presently observed dwarfs (``survived ones'')
and thus result in the observed kinematical differences.
In the present paper, we consider that
the observed line-of-sight velocity dispersions 
of UCDs and dwarf ellipticals (dEs)  can reflect the intrinsic differences
in orbital properties between these two dwarf galaxy populations.
We thus investigate kinematical properties of different galaxy populations
with different $r_{\rm p}$ and  $e$ in the Fornax cluster model.

\subsection{Orbital evolution of galaxy populations in the Fornax cluster}

We investigate orbital evolution of ``galaxies'' represented
by test particles in fixed gravitational potentials reasonable
for clusters of galaxies.
We adopt orbital evolution models similar to those used by 
Bekki et al. (2003a) for better understanding the origin of the spatial
distributions of UCDs in clusters and groups.
We consider that  particles  with the total number
of $N$ orbit a cluster dominated by 
a massive dark matter halo with the total mass of $M_{\rm cl}$. 
To give our model a realistic radial density profile for
the dark matter halo of the cluster,
we base it on the predictions from the standard cold
dark matter cosmogony (Navarro, Frenk, \& White 1996, hereafter NFW).
The NFW profile is described as:
\begin{equation}
{\rho}(r)=\frac{\rho_{0}}{(r/r_{\rm s})(1+r/r_{\rm s})^2},
\end{equation}
where $r$,  $\rho_{0}$,  and $r_{\rm s}$ are the distance from the center
of the cluster, the central density, and the scale-length of the dark halo,
respectively.

We mainly investigate the ``Fornax'' model with 
$M_{\rm cl}=7.0 \times 10^{13} \rm M_{\odot}$,
$r_{\rm s}$ of 83\,kpc, and ``c'' parameter of 12.8,
because we are interested in the observed kinematics of UCDs and other
galaxy populations in the Fornax.
The adopted set of parameters can be  consistent with the
X-ray observations of Jones et al. (1997) and
the mass estimation of the Fornax cluster derived 
from  kinematics of the cluster
galaxies
by Drinkwater et al. (2001).
We do not consider gravitational influences of cluster member galaxies
including NGC 1399 in the Fornax model, 
because the influences can be very minor owing to
the total mass of the galaxies much smaller than that of  
the background dark matter halo.

For the projected radial number distribution of particles
($\Sigma (R)$) within the  cluster,
we adopt a King profile with a core radius $\sim 0.6$ times
smaller than the cluster scale radius, $r_{\rm s}$, 
in each cluster model (Adami et al. 1998).
In order to derive the three dimensional (3D) density field
(${\rho}(r)$, where $r$ is the distance from the
center of the cluster)
from $\Sigma (R)$, we can use the following formula
(Binney \& Tremaine  1987):
\begin{equation}
\rho (r)
 = -\frac{1}{\pi} {\int}_{r}^{\infty}
 \frac{d\Sigma(R)}{dR}
 \frac{dR}{\sqrt{R^2-r^2}}.
\end{equation}
We numerically estimate the spherical symmetric $\rho (r)$ profile
for a given $\Sigma (R)$ in a model.

\begin{table}
\centering
\begin{minipage}{85mm}
\caption{Model parameters}
\begin{tabular}{cccc}
Model no.  
& {$e_{\rm m,0}$
\footnote{The initial mean of
orbital eccentricities ($e$) for particles.}}
& {${\sigma}_{0}$
\footnote{The initial 
dispersion in $e$ for particles.}}
& comments \\ 
M1 & 0.6  & 0.2  & standard model \\
M2 & 0.5  & 0.2  & more circular orbits\\
M3 & 0.7  & 0.2  & more eccentric orbits \\
M4 & 0.6  & 0.1  & \\
M5 & 0.6  & 0.3  & \\
\end{tabular}
\end{minipage}
\end{table}

Each particle is given an  initial orbital eccentricity ($e$)
and a pericenter distance ($r_{\rm p}$) for a given cluster potential.
We give each particle $e$  so that
the $e$-distribution for $N$ particles has a Gaussian with
the  mean $e$ value and the dispersion of $e$ 
being $e_{\rm m, 0}$ and ${\sigma}_{0}$, respectively. 
Since we adopt spherical distributions of particles,
we do not introduce initial anisotropy in velocity dispersion
for the $x$, $y$, and $z$ directions: the velocity ellipsoid
of particles can be  anisotropic in radial ($r$) direction only.
In this calculation, we consider 
that the mean orbital eccentricity of galaxies in a cluster 
($e_{\rm m,0}$) is 0.6, which is
consistent with recent high-resolution 
cosmological simulations (Ghigna et al. 1998).
We however investigate different five sets 
of models with $0.5\le e_{\rm m, 0} \le 0.7$
and $0.1 \le {\sigma}_{0} \le 0.3$.

We investigate radial profiles of (projected) line-of-sight
velocity dispersions (${\sigma}_{\rm los}(R)$)
and velocity dispersions
within 200 kpc (${\sigma}_{m}$) for 
``selected particles'' and all ones.  We investigate ${\sigma}_{m}$ within 200 kpc,
mainly because all UCDs are located within the central 200 kpc
of the Fornax (e.g., Drinkwater et al. 2003; Gregg et al. 2007).
We consider that particles with different $r_{\rm p}$  and $e$
can have different kinematics,  and accordingly we investigate
${\sigma}_{\rm los}(R)$ and ${\sigma}_{m}$
for ``selected particles'' with $r_{\rm p}$ ($e$)  being within 
a certain parameter range.
We introduce two key parameters; (1) $r_{\rm th}$  which
$r_{\rm p}$ of selected particles are smaller than and (2)  $e_{\rm th}$
which $e$ of the particles are larger than.
We consider that  $r_{\rm th}$ and $e_{\rm th}$ 
are key, because our previous simulations of UCD formation based
on the threshing scenario suggest that these two are important
for UCD formation.

We investigate the models with  25 kpc $\le r_{\rm th} \le $ 175 kpc
and $0.5 \le e_{\rm th} \le 0.7$  for a given $e_{\rm m,0}$ and
${\sigma}_0$. We mainly show the results of the ``standard'' model
(M1)
with $e_{\rm m,0}=0.6$ and ${\sigma}_0=0.2$, because the adopted
parameters are the most reasonable in the present study.
The parameter values of other four models (M2, M3, M4, and M5)
are given in the Table 1. 
Since most UCDs are located at  50 kpc$ \le R \le $200 kpc
in the Fornax cluster (Gregg et al. 2007),
we estimate  ${\sigma}_{m}$ for particles with 50 kpc$ \le R \le $200 kpc.  
For conveniences, 
${\sigma}_{m}$ for selected and all particles are referred
to as ${\sigma}_{\rm m}$ and ${\sigma}_{\rm m,all}$,
respectively,  in the present study.

$N$ is set to be $10^4$ for all models so that
the error bar in each $i$-th radial ($R$) bin due to the smaller
number of particles (i.e., $\sim 1/\sqrt{N_{i}}$) can be as small as 
$\sim 20$ km s$^{-1}$.  For models with $N$ as large as $10^4$, 
we can more clearly
see differences in kinematics between selected and all particles
owing to the smaller error bars,
though $N \sim10^4$ could be significantly larger than a reasonable
number of galaxy populations and intracluster stellar systems
like UCDs.

Thus  we adopt somewhat idealized orbital evolution models for
galaxies in the cluster and thereby investigate kinematical
differences between different orbital populations. 
The results presented in this study will be useful and helpful
not only in interpreting observational results on UCD kinematics
but also in understanding  
the results of much more sophisticated simulations with UCD formation
models in our future numerical studies. 
Orbital properties, ages and metallicities, spatial distributions
of UCDs derived in our cosmological simulations
will be given in our future papers (Bekki \& Yahagi 2007, in preparation).
Gregg et al. (2007) have  revealed that the spatial distribution
of UCDs in the Fornax
is quite compact and appears to be significantly flattened.
We will also discuss the origin of the observed  flattened
spatial distribution 
in the context of formation of hosts galaxies of UCDs at
very high redshifts ($z>10$)
based on the results of these future  cosmological simulations.

\begin{figure}
\psfig{file=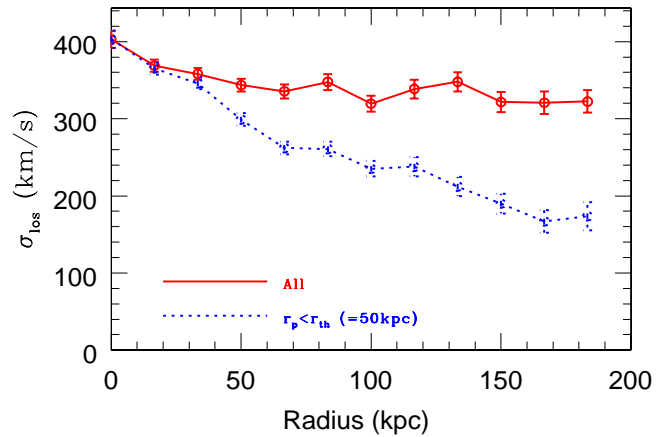,width=8.5cm}
\caption{
Dependences of line-of-sight velocity dispersions (${\sigma}_{\rm los}$)
on projected radii ($R$) for all particles (solid, red) and those with
$r_{\rm p}$  smaller
than a threshold radius ($r_{\rm th}$) of 50 kpc (dotted, blue). 
Note that ${\sigma}_{\rm los}(R)$ is significantly steeper
and systematically lower in the model with selected particles
with $r_{\rm p}<r_{\rm th}$.
}
\label{Figure. 1}
\end{figure}

\begin{figure}
\psfig{file=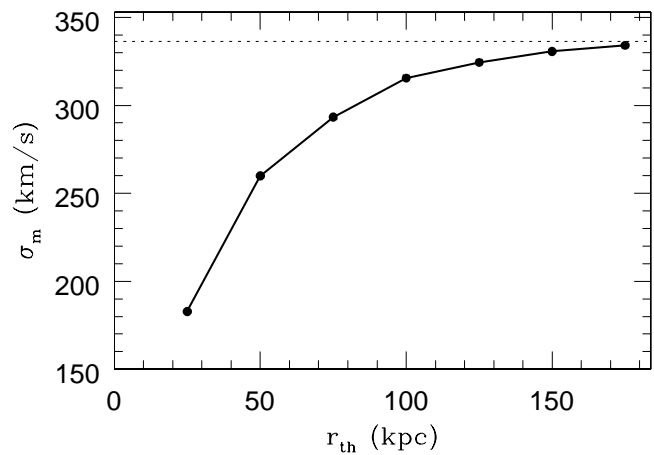,width=8.5cm}
\caption{
The dependence of line-of-sight velocity dispersion of particles within the
central 200 kpc (${\sigma}_{\rm m}$) 
on $r_{\rm th}$ (=25, 50, 75, 100, 125, 150, and 175 km s$^{-1}$)
in the standard model (solid).
${\sigma}_{\rm m}$ of all particles is shown by a dotted line
for comparison.
}
\label{Figure. 2}
\end{figure}

\begin{figure}
\psfig{file=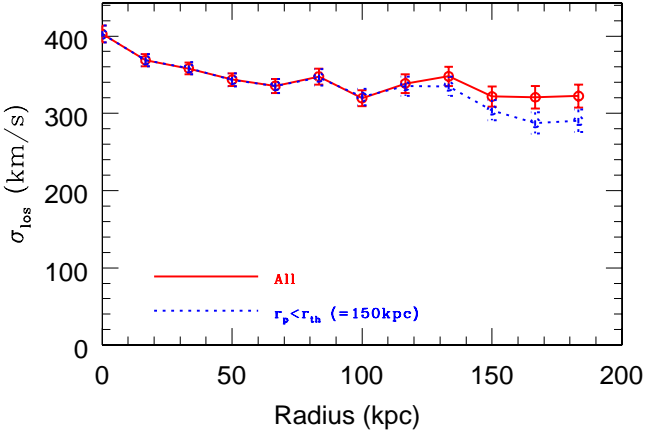,width=8.5cm}
\caption{
The same as Figure 1 but for the model with $r_{\rm th}$=150 kpc.
}
\label{Figure. 3}
\end{figure}

\section{Results}

\subsection{The standard model}

Figure 1 clearly shows that ${\sigma}_{\rm los}(R)$
is significantly steeper and systematically
smaller in selected particles with
$r_{\rm p}<r_{\rm th}$  than in all ones for the central 200
kpc of the standard model with $r_{\rm th}=50$ kpc.
The steeper  ${\sigma}_{\rm los}(R)$ profile
in the particles with $r_{\rm p}<50$ kpc
is due to a large number of more eccentric orbits
(i.e., a radially anisotropic velocity ellipsoid).
The difference in ${\sigma}_{\rm los}(R)$ becomes 
progressively larger with larger $R$, which reflects
the fact that the outer selected particles
need to have more eccentric orbits to have $r_{\rm p}<50$ kpc.
The velocity dispersions within the central 200 kpc
for all particles (${\sigma}_{\rm m,all}$) and selected
ones (${\sigma}_{\rm m}$) are 
336 km s$^{-1}$  and 260  km s$^{-1}$, respectively,
for this standard model.

Figure 2 shows that  ${\sigma}_{\rm m}$ depends on
$r_{\rm p}<r_{\rm th}$ (25 kpc$\le r_{\rm th} \le$175 kpc) such that
it is smaller for smaller $r_{\rm th}$ in the standard model:
${\sigma}_{\rm m}/{\sigma}_{\rm m,all}$ is 0.54 for $r_{\rm th}=25$ kpc
and 0.99 for $r_{\rm th}=175$ kpc.
The physical reason for this dependence is that
the model with smaller  $r_{\rm th}$
has a larger number of particles with more eccentric orbits
in the outer part of the simulated region (50 kpc$ \le R \le $200 kpc).
Figure 3 shows that the standard model with $r_{\rm th}=150$ kpc
shows a very minor difference in  ${\sigma}_{\rm los}(R)$
between all and selected particles, which
confirms that  $r_{\rm th}$ is a key parameter for kinematics
in the present models for a given $e_{\rm m,0}$ and ${\sigma}_0$.

Figure 4 shows that ${\sigma}_{\rm m}$
depends very weakly on $e_{\rm th}$ for $0.5 \le e_{\rm th} \le 0.7$ 
in the sense that ${\sigma}_{\rm m}$ is smaller for larger
$e_{\rm th}$ (i.e., more eccentric orbits).
It is confirmed that ${\sigma}_{\rm los}(R)$ is steeper
in the selected particles with $e > e_{\rm th}$
than in all ones for the central region ($R<20$ kpc).
These results strongly suggest that $r_{\rm th}$ is
a much more important parameter than  $e_{\rm th}$ 
for velocity dispersions in the standard model.
These results in Figures $1-4$  imply that
the observed lower velocity dispersions of UCD populations
in clusters of galaxies have physical origins: the lower dispersions
are  unlikely to result from some uncertainties
in observational
methods to derive ${\sigma}_{\rm m}$ for a smaller number of
UCDs.

\begin{figure}
\psfig{file=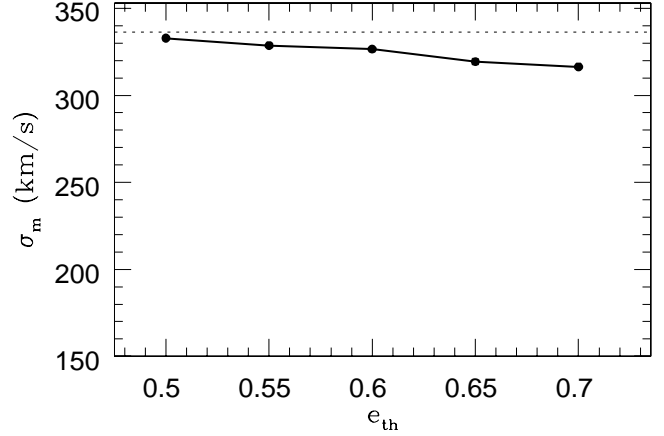,width=8.5cm}
\caption{
The dependence of 
${\sigma}_{\rm m}$ 
on $e_{\rm th}$ (=0.5, 0.55, 0.6, 0.65, and 0.70)
in the standard model (solid).
${\sigma}_{\rm m}$ of all particles is shown by a dotted line
for comparison.
}
\label{Figure. 4}
\end{figure}

\subsection{Parameter dependences}
The three important
parameter dependences of ${\sigma}_{\rm m}$  and
${\sigma}_{\rm los}(R)$ on initial orbital properties
of particles in cluster models 
(i.e.,  $e_{\rm m,0}$ and ${\sigma}_{\rm 0}$)
are briefly summarized as follows.
Firstly,  the dependence of ${\sigma}_{\rm m}$  
on $r_{\rm th}$ derived in the standard model
is seen in  the models with different 
$e_{\rm m,0}$ and ${\sigma}_{\rm 0}$.
Figure 5 clearly shows that irrespective of 
$e_{\rm m,0}$ and ${\sigma}_{\rm 0}$,
${\sigma}_{\rm m}$ is smaller for smaller $r_{\rm th}$,
which confirms that $r_{\rm th}$ is a key parameter 
for kinematical properties of galaxy populations
in clusters.

Secondly, ${\sigma}_{\rm los}(R)$ is significantly
steeper and systematically smaller in the selected
particles with  $r<r_{\rm th}$ than in all ones for models with
different $e_{\rm m,0}$ and ${\sigma}_{\rm 0}$,
though the details of the profiles depends slightly on
$e_{\rm m,0}$ and ${\sigma}_{\rm 0}$. 
Thirdly dependences of ${\sigma}_{\rm m}$
and  ${\sigma}_{\rm los}(R)$  on $e_{\rm th}$ 
are much weaker than those on  $r_{\rm th}$ for different models,
which suggests that $r_{\rm th}$ is a more important
parameter for kinematics of galaxy populations.
These results for different models 
are thus all consistent with 
those derived for the standard model.

\begin{figure}
\psfig{file=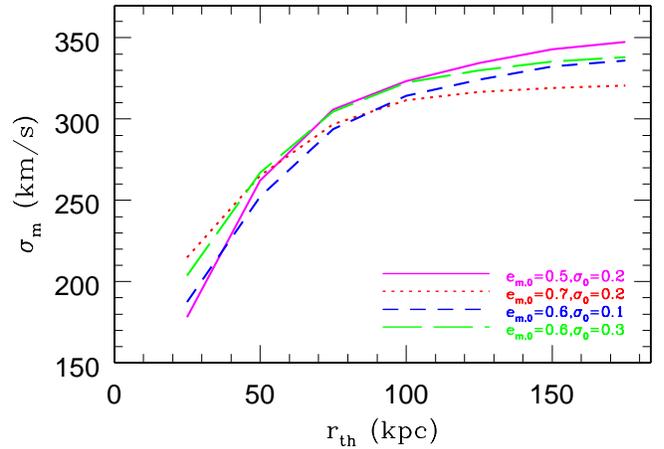,width=8.5cm}
\caption{
The same as Figure 2 but for different models: 
M2 with $e_{\rm m,0}=0.5$ and ${\sigma}_{0}=0.2$ (solid, magenta),
M3 with $e_{\rm m,0}=0.7$ and ${\sigma}_{0}=0.2$ (dotted, red),
M4 with $e_{\rm m,0}=0.6$ and ${\sigma}_{0}=0.1$ (dashed, blue), and
M5 with $e_{\rm m,0}=0.6$ and ${\sigma}_{0}=0.3$ (long-dashed, green).
}
\label{Figure. 5}
\end{figure}

\section{Discussions}

\subsection{UCDs originate  from 
destroyed building blocks of clusters ?}

The present study has first shown that
galaxies with smaller $r_{\rm p}$ ($< 50$ kpc)
can show smaller velocity dispersions in the central
200 kpc of the  cluster models.
This result implies that the observed
lower velocity dispersions  in 
the Fornax and the Virgo can be due to
UCDs having smaller $r_{\rm p}$ ($< 50$ kpc)
in comparison with other galaxy populations in these clusters.
The result also suggests that
the observed lower dispersions are consistent with 
the galaxy threshing scenario in which only  galaxies with very small 
$r_{\rm p}$  can become UCDs after the destruction of
the galaxies via strong cluster tidal fields (Bekki et al. 2001; 2003a).
Thus the observed lower velocity dispersions of UCD populations
can provide a clue to the formation processes of UCDs in clusters.

It should be however stressed that the observed lower dispersions
of UCDs
can not rule out the alternative scenario 
by Fellhauer \&  Kroupa (2002)
in which UCDs can be formed by merging of numerous smaller clusters
in starbursting  galaxy mergers (``cluster merger scenario''):
if young UCDs formed in galaxy mergers 
are tidally stripped to have more radial orbits and smaller
$r_{\rm p}$ (for some physical reasons) 
in the central
regions  of clusters,
the observed lower dispersions would be explained equally.
Since galaxy merging is highly unlikely 
after the formation of clusters at low redshifts 
(e.g., Ghigna et al. 1998),  galaxy merging that can form
UCDs probably needs to occur in the very early stages of 
clusters of galaxies.
Thus, although the observed lower velocity dispersions of UCDs
are consistent with the threshing scenario,
they still appear not to be  evidences strong enough  
to determine which of the above two scenarios is  more
reasonable.

\subsection{Possible kinematical differences between UCDs and ICGCs}

Recent observational studies of GCs
in clusters of galaxies have suggested that
there can be a population of GCs that are
bounded by cluster gravitational potentials
rather than those of cluster member
galaxies (e.g., West et al. 1995; Bassino et al.  2003; 
Jord\'an et al. 2003; 	
Tamura et al. 2006; Bergond et al. 2007;
Williams et al. 2007).
Previous numerical simulations showed that
these intracluster GCs (ICGCs) can be formed from
tidal stripping of GCs that are located in the outer parts
of galaxies orbiting clusters of galaxies (e.g., Bekki et al. 2003b).
Given that the UCDs can be the nuclei of destroyed (thus defunct)
galaxies,  spatial distributions and kinematics of ICGCs and UCDs
in a cluster can be significantly different.

Bekki \& Yahagi (2006) suggested that ICGCs can show more isotropic
line-of-sight velocity dispersions, which can be in 
a contrast with the possibly (radially) anisotropic ones
of UCDs suggested by the galaxy threshing scenario (Bekki et al. 2003).
Although observational studies on structural  and kinematical properties
of ICGCs in nearby clusters have just started (e.g., Tamura et al. 2006;
Bergond et al. 2007),
these dynamical properties for relatively faint ICGCs have not yet been
derived for the {\it entire} regions of the clusters.
It is therefore observationally unclear whether spatial distributions
and kinematics within clusters are different between UCDs and ICGCs. 
Future observations on the differences in ${\sigma}_{\rm los}(R)$ 
between UCDs and ICGCs may well help us to understand
the differences in formation processes (e.g., threshing vs stripping)
between these two intracluster  populations.

\section{Conclusions}

We have mainly 
investigated orbital evolution of galaxies with different
orbital parameters (i.e., $e$ and $r_{\rm p}$) in 
the central region ($R<200$ kpc) of the Fornax
cluster model in order to understand the origin of the observed
lower velocity dispersion of UCDs in the Fornax.
We have introduced two key  parameters  $r_{\rm th}$ and $e_{\rm th}$
and thereby
investigated ${\sigma}_{\rm m}$
and ${\sigma}_{\rm los}(R)$ 
for galaxies with $r_{\rm p} < r_{\rm th}$ 
(or with $e < e_{\rm th}$) for a given
$e_{\rm m,0}$ and ${\sigma}_{0}$.
We have particularly focused on the kinematical differences
between all galaxies (e.g., ${\sigma}_{\rm m,all}$)
and  those selected with $r_{\rm p}<r_{\rm th}$ (e.g., ${\sigma}_{\rm m}$).
We summarise our principle results as
follows.

(1) ${\sigma}_{\rm los}(R)$ is steeper in selected galaxies
with  $r_{\rm p}<r_{\rm th}$ than in all ones 
for the standard model with $e_{\rm m,0}=0.6$ and ${\sigma}_{0}=0.2$.  
${\sigma}_{\rm m}$ is significantly smaller than  ${\sigma}_{\rm m,all}$
in the standard model and the differences between ${\sigma}_{\rm m}$
and ${\sigma}_{\rm m,all}$ depend on $r_{\rm th}$ in
such a way that they can be larger for smaller $r_{\rm th}$
for 25 kpc$\le r_{\rm th} \le$175 kpc.

(2) ${\sigma}_{\rm los}(R)$ is steeper in selected galaxies
with  $e_{\rm p}>e_{\rm th}$ than in all ones 
for the standard model. Although ${\sigma}_{\rm m}$ 
of the selected galaxies with  $e_{\rm p}>e_{\rm th}$ 
is slightly smaller than ${\sigma}_{\rm m,all}$ 
in the standard model,
the difference between the two can be quite small ($\sim 20$ km s$^{-1}$)
for $0.5 \le e_{\rm th} \le 0.7$.

(3) The above results derived for the standard model  
are seen in the models with different $e_{\rm m,0}$ and ${\sigma}_{0}$,
which implies that lower velocity dispersions of UCDs can be 
due to their smaller $r_{\rm p}$ in comparison with other galaxy
populations in the Fornax. 
Thus the observed kinematical differences in UCDs and other dwarf
galaxy populations are due to differences in 
intrinsic orbital properties between these galaxy populations
in the Fornax cluster.

Although the observed significantly flattened spatial distribution 
of UCDs in the Fornax cluster
can potentially have important implications on dwarf galaxy formation
in the cluster  (Gregg et al. 2007),
we did not discuss the origin of the distribution.
We plan to discuss the observed distribution 
in the context of the picture that
UCDs originate from galaxies embedded in
dark matter halos already virialized at  $z>10$ 
in the proto-cluster
environment (Bekki \& Yahagi 2007, in preparation).

\section*{Acknowledgments}
We are  grateful to the anonymous referee for valuable comments,
which contribute to improve the present paper.
K.B. acknowledges the financial support of the Australian Research
Council throughout the course of this work.


\begin{thebibliography}{99}

\bibitem[]{}
Adami, C., Mazure, A., Katgert, P.,  Biviano, A.,
1998, A\&A, 336, 63


\bibitem[]{}
Bassino, L. P., Cellone, S. A., Forte, J. C.,   Dirsch, B.
2003, A\&A, 399, 489


\bibitem[]{}
Bekki, K., Couch, W. J., Drinkwater, M. J.,
2001, ApJL, 552, 105 

\bibitem[]{}
Bekki, K., Couch, W. J., Drinkwater, M. J.,  Shioya, Y. 2003a,  
MNRAS, 344, 399

\bibitem[]{} 
Bekki, K., Forbes, D. A.,  Beasley, M. A.,   Couch, W. J.,
2003b, MNRAS, 344, 1334


\bibitem[]{}
Bekki, K., Yahagi, H., 2006, MNRAS, 371, 1019
	
\bibitem[]{}
Bekki, K., Couch, W. J.,  Shioya, Y., 2006, ApJL, 642, 133

\bibitem[]{}
Bekki, K., Yahagi, H.,  Forbes, D. A., 2007, preprint (astro-ph/0702088)


\bibitem[]{}
Bergond, G. et al., 2007, accepted by  A\&A Letters (astro-ph/0701378)

\bibitem[Binney \& Tremaine 1987]{bi87}
Binney, J.,  Tremaine, S., 1987 in Galactic Dynamics.


\bibitem[]{}
De Propris, Phillipps, S.,
Drinkwater, M. J.,
Gregg, M. D.,
Jones, J. B.,
Evstigneeva, E., 
Bekki, K.,
2005, ApJL, 623, 105


\bibitem[]{}
Dirsch, B. et al., 2004, AJ, 127, 2114

\bibitem[Drinkwater  et al.  2000a]{dr00a}
Drinkwater, M. J., Phillipps, S.,  Jones, J. B., Gregg, M. D.,   Deady, J. H., 
Davies, J. I., Parker, Q. A., Sadler, E. M.,  Smith, R. M. 2000a,
A\&A, 355, 900 

\bibitem[Drinkwater  et al.  2000b]{dr00b}
Drinkwater, M. J., Jones, J. B., Gregg, M. D.,  Phillipps, S.  2000b,
PASA, 17, 227 

\bibitem[]{}
Drinkwater, M. J., Gregg, M. D., Colless, M. 2001, ApJL, 548, 139

\bibitem[]{}
Drinkwater, M. J., Gregg, M. D.,  Hilker, M., Bekki, K., Couch, W. J.,
Ferguson, J. B.,  Jones, J. B.,   Phillipps, S. 2003, Nat, 423, 519

\bibitem[]{}
Evstigneeva, E. A., Gregg, M. D.,  Drinkwater, M. D., Hilker, M.,
2007, accepted for AJ, (astro-ph/0612483)

\bibitem[]{}
Fellhauer, M., Kroupa, P. 2002, MNRAS, 330, 642

\bibitem[]{}
Firth, P. et al., 2006,
in the Globular Clusters to Guides to
Galaxies (astro-ph/0606041)

\bibitem[]{}
Ghigna, S., Moore, B., Governato, F., Lake, G.,
Quinn, T., Stadel, J. 1998, MNRAS, 300, 146

\bibitem[]{}
Gregg, M. D. et al, 2007, submitted to AJ


\bibitem[]{}
Ha\c segan et al., 2005, ApJ, 627, 203


\bibitem[]{}
Hilker, M., Infante, L.,   Richtler, T. 1999, A\&AS, 138, 55

\bibitem[]{}
Hilker, M.,  Baumgardt, H.,  Infante, L.,  Drinkwater, M. D., Evstigneeva, E. A.,
Gregg. M. D., 2007, accepted for A\&A, (astro-ph/0612484)


\bibitem[]{}
Jones, C., Stern, C., Forman, W., Breen, J., David, L., Tucker, W., 
Franx, M. 1997, ApJ, 482, 143


\bibitem[]{}
Jones, J. B. et al., 
2006, AJ, 131, 312


\bibitem[]{} 
Jord\'an, A.,  West, M. J.,
C\^ote,~P.,   Marzke, R. O., 2003, AJ, 125, 1642 



\bibitem[]{}
Karick, A. M., Gregg, M. D., Drinkwater, M. J., Hilker, M.,
Firth, P., 2006,  in the Globular Clusters to Guides to
Galaxies (astro-ph/0605515)
 
\bibitem[]{}
Kilborn, V. A. et al., 2005, PASA, 22, 326

%\bibitem[King  1962]{kin62}
%King, I. R. 1962, AJ, 67, 471

\bibitem[]{}
Mieske, S., Hilker, M., Infante, L., 2002, A\&A, 383, 823


\bibitem[]{}
Mieske, S., Hilker, M., Infante, L., 2004, A\&A, 418, 445

\bibitem[]{}
Mieske, S., Hilker, M., Infante, L., Jord\'an, A., 2006, AJ, 131, 2422 

\bibitem[Navarro et al. 1996]{na96}
Navarro, J. F., Frenk, C. S.,  White, S. D. M.
1996, ApJ, 462, 563

\bibitem[]{}
Phillipps, S., Drinkwater, M. J., Gregg, M. D., Jones, J. B.,
2001, ApJ, 560, 201


\bibitem[]{} 
Tamura, N., Sharples, R. M., Arimoto, N., Onodera, M., Ohta, K., Yamada, Y.,
2006, MNRAS, 373, 601

\bibitem[]{} 
West, M. J.,  C\^ote,~P.,  Jones, C.,  Forman, W.,
Marzke, R. O.,  1995, ApJ, 453, L77

\bibitem[]{}
Williams, B. et al., 2007, ApJ, 654, 835

\bibitem[]{}
Yahagi, H.,  Bekki, K., 2005, MNRAS, L364, 86

\end{thebibliography}
\end{document}